\documentclass[12pt,hyper]{JHEP3}
\usepackage{amssymb}
\usepackage{mathrsfs} %script fonts 
\usepackage{graphicx}
\usepackage{bm}% bold math
\usepackage{setstack}
\usepackage{wasysym}
\def\A{{\mathscr{A}}}
\def\N{{\mathscr{N}}}
\title{Entropy bounds for uncollapsed rotating bodies}

\author{
Gabriel Abreu \textrm{and} Matt Visser\\
School of Mathematics, Statistics, and Operations Research\\
Victoria University of Wellington \\
PO Box 600, Wellington 6140, 
New Zealand\\
E-mail: \email{gabriel.abreu@msor.vuw.ac.nz, matt.visser@msor.vuw.ac.nz}

}

\abstract{
Entropy bounds in black hole physics,  based on a wide variety of different approaches, have had a long and distinguished history. Recently the current authors have turned attention to uncollapsed systems and obtained a robust  entropy bound for uncollapsed static spherically symmetric configurations.
In the current article we extend this bound to rotating systems. This extension is less simple than one might at first suppose.

Purely classically, (using only classical general relativity and basic thermodynamics), it is possible to show that the entropy of uncollapsed matter inside a region enclosed by a surface of area $\A$ is bounded from above by
\[
S \leq {||\vec \kappa||_{\mathrm{max(surface)}} \; \A \over 4\pi \; T_\infty}.
\]
Here $\vec \kappa\,$ is a suitably defined surface gravity.  By appealing to the Unruh effect, which is our only invocation of quantum physics, we argue that for a suitable class of fiducial observers there is a (quantitatively weak but qualitatively robust) lower bound on the temperature (as measured at spatial infinity)
\[
T_{\infty}  \geq \max\nolimits_{_\mathrm{FIDOs}} \left\{ {||\vec\kappa || \over2\pi} \right\} .
\]
Thus, using only classical general relativity, basic thermodynamics, and the Unruh effect, we are able to argue that for uncollapsed matter
\[
S \leq {1\over2} \; \A.
\]
  }

\begin{document}
%-------------------------------------------------------------------------

%----------------------------------------------------
\def\implies{\Rightarrow}
%----------------------------------------------------
\renewcommand{\sun}{\ensuremath{\odot}}%
\def\ie{{\emph{i.e.}}}
\def\Barcelo{Barcel\'o}
\def\d{{\mathrm{d}}}
\def\etc{\emph{etc}}
\def\A{{\mathscr{A}}}
%----------------------------------------------------
\section{Introduction}
%----------------------------------------------------

Consider a rotating blob of matter that has not collapsed to a black hole. Can we nevertheless place a robust bound on its entropy using fundamental physics, without resorting to black hole physics (generalized second law~\cite{Bekenstein-gsl}, holographic bound~\cite{holographic}, horizon entanglement entropy~\cite{Srednicki}) or the Bekenstein bound~\cite{Bekenstein-bound}? 

In the case of non-rotating static configurations this was recently answered in the affirmative~\cite{Abreu1, Abreu2} when the present authors derived a bound based only on classical general relativity, basic thermodynamics, and the Unruh effect~\cite{Unruh}, to the effect that (entropy) $\leq {1\over2}$ (area).  The ${1\over2}$ is not a typo --- ultimately one has  ${1\over2}$  (rather than the  ${1\over4}$ one might naively expect based on black hole physics) simply because the matter has \emph{not} collapsed to a black hole~\cite{Abreu1, Abreu2}.
We shall now extend this result to rotating blobs of uncollapsed matter --- the derivation is slightly tricker than one might at first suppose, and the logic flow has to be somewhat re-ordered, but ultimately the basic result is the same: (entropy) $\leq {1\over2}$ (area).

To start the discussion we appeal to the \emph{ordinary} second law to note that the entropy of the rotating blob is less than what it will be once the blob has settled down to complete mechanical and thermodynamic equilibrium. So we might as well restrict attention to equilibrium configurations. But equilibrium configurations in general relativity have three absolutely crucial properties. They are:
\begin{itemize}
\item Stationarity.
\item Azimuthal symmetry.
\item Rigid rotation.
\end{itemize}
Stationarity basically means time-independence, a basic requirement of equilibrium configurations.  Turning to the remaining two conditions: Physically, if the blob is not azimuthally symmetric, but is rotating, then it will emit gravitational radiation, thereby losing energy, so it cannot be in equilibrium.  Thus equilibrium in general relativity implies the standard result that there are two Killing vectors, one timelike and one spacelike. (See for example~\cite{Hartle, Israel, Wai-Mo}.)  Finally, if the blob is a solid, then ``rigid rotation'' is automatic. For a fluid the physics argument goes as follows: If the fluid blob is not rigidly rotating then velocity gradients imply shear, and shear implies friction, so the blob is losing energy, and cannot be in equilibrium. Thus in general relativity an equilibrium body cannot support any shear, and this will be our definition of rigid rotation. (See for example~\cite{Hartle, Israel, Wai-Mo}).  Since these are quite standard results we will simply use them and not further belabour the point. 

The goal now is, within this particular framework, to derive robust bounds on the entropy content of the rotating blob.  We shall first derive a purely classical upper bound on the entropy, using only classical general relativity and basic thermodynamics.  This purely classical bound, because the terms being neglected are comparable to the terms being retained, is quantitatively relatively strong. 

We then make our only appeal to quantum physics, using the Unruh effect to argue for a (quantitatively weak but qualitatively robust) semiclassical lower bound on the temperature based on the 4-acceleration of a suitably defined class of fiducial observers [FIDOs]. Combining these results we argue that for uncollapsed matter there is a bound: (entropy) $\leq {1\over2}$  (area).   Because of the relative weakness of the bound on the temperature, this semiclassical entropy bound is also quantitatively relatively weak --- unless the object of interest is extremely compact. Such bounds are of particular interest in view of recent speculations regarding monsters~\cite{monster1, monster2, monster3}/ gravastars~\cite{gravastar1, gravastar2, gravastar3, gravastar4}, black stars~\cite{black-star1, black-star2, black-star3, black-star4}, or quasi-black holes~\cite{qbh1, qbh2}.

%------------------------------------------------------------------------------------------------------------------------------------------
\section{Metric}
%------------------------------------------------------------------------------------------------------------------------------------------

In view of stationarity and azimuthal symmetry we can choose coordinates such that the metric takes the form
\begin{equation}
\d s^2 = g_{tt} \d t^2 + 2 g_{t\phi} \d t \d\phi + g_{\phi\phi} \d \phi^2 + g_{rr} \d r^2 + g_{\theta\theta} \d \theta^2, 
\end{equation}
which is better written as
\begin{equation}
\d s^2 = -N^2 \d t^2 + g_{\phi\phi} (\d \phi -\varpi\d t)^2 + g_{rr} \d r^2 + g_{\theta\theta} \d \theta^2.
\end{equation}
Note that the metric components are independent of $t$ and $\phi$. The labels $r$ and $\theta$ for the remaining two coordinates are completely conventional and these coordinates can be shuffled around at will (as long as one does so in a $t$ and $\phi$ independent manner).  Note that we use  $-\varpi$ to denote the ratio of metric components $g_{t\phi}/g_{\phi\phi}$;  the symbol $\omega$ will be reserved for the vorticity of a certain timelike congruence we shall subsequently encounter. 
This ADM decomposition for the stationary axially symmetric spacetime implies
\begin{equation}
g_{ab} 
= \left[
\begin{array}{cccc}
g_{tt}  & g_{t\phi}  & 0 & 0  \\
g_{t\phi}  & g_{\phi\phi}  & 0 & 0  \\
0  & 0  & g_{rr} & 0\\
0 & 0 & 0 & g_{\theta\theta}  
\end{array}
\right]
=
\left[
\begin{array}{cccc}
- [N^2- g_{\phi\phi} \varpi^2 ]& -g_{\phi\phi} \varpi & 0 & 0  \\
  -g_{\phi\phi} \varpi  & g_{\phi\phi}  & 0 & 0  \\
0  & 0  & g_{rr} & 0\\
0 & 0 & 0 & g_{\theta\theta}  
\end{array}
\right],
\end{equation}
and
\begin{equation}
g^{ab} 
= \left[
\begin{array}{cccc}
g^{tt}  & g^{t\phi}  & 0 & 0  \\
g^{t\phi}  & g^{\phi\phi}  & 0 & 0  \\
0  & 0  & g^{rr} & 0\\
0 & 0 & 0 & g^{\theta\theta}  
\end{array}
\right]
=
\left[
\begin{array}{cccc}
- {1/N^2}& -\varpi/N^2 & 0 & 0  \\
 -\varpi/N^2  & \; \;1/g_{\phi\phi}- \varpi^2/N^2 \; & 0 & 0  \\
0  & 0  & 1/g_{rr} & 0\\
0 & 0 & 0 & 1/g_{\theta\theta}  
\end{array}
\right].
\end{equation}
Note also that on the rotation axis we have $g_{\phi\phi}\to0$.

%------------------------------------------------------------------------------------------------------------------------------------------
\section{Matter}
%------------------------------------------------------------------------------------------------------------------------------------------

The two natural Killing vectors are the timelike Killing vector
\begin{equation}
(k_T)^a = (\partial_t)^a = (1,0,0,0),
\end{equation} 
and the azimuthal Killing vector 
\begin{equation}
(k_\Phi)^a = (\partial_\phi)^a = (0,1,0,0).
\end{equation} 
Assuming the matter is a perfect fluid, with the SET taking the form
\begin{equation}
T^{ab} = (\rho+p) u^a u^b + p g^{ab},
\end{equation}
then internal equilibrium within the fluid ball demands~\cite{Hartle, Israel, Wai-Mo}
\begin{equation}
u^a \propto (k_T)^a + \Omega \; (k_\Phi)^a,
\end{equation}
where $\Omega$ is constant throughout the fluid. That is, there is a \emph{comoving} Killing vector
\begin{equation}
(k_C)^a  = (k_T)^a + \Omega (k_\Phi)^a =   (\partial_t)^a + \Omega (\partial_\phi)^a = (1,\Omega,0,0),
\end{equation} 
such that
\begin{equation}
u_C^a =   {(k_C)^a\over ||k_C||}.
\end{equation}
It is trivial to see that the shear of this 4-velocity $u_C$ is zero. The converse is slightly tedious but completely standard~\cite{Hartle, Israel, Wai-Mo}.
A further nontrivial and potentially useful observation is that   $\Omega>\varpi$ is guaranteed throughout the interior of the blob via the so-called ``$r$-mode instability'' --- if at any point internal to the fluid ball we have $\varpi>\Omega$ then the fluid blob cannot be in internal equilibrium~\cite{r-mode1, r-mode2, r-mode3}.  Finally, observe that the comoving 4-velocity is spacelike far away from the axis of rotation --- this is perfectly standard, and in particular implies a physical bound on the spatial size of the rotating blob --- in the sense that the surface of the blob must have 3-velocity slower than the speed of light.

%------------------------------------------------------------------------------------------------------------------------------------------
\section{Thermodynamic equilibrium}
%------------------------------------------------------------------------------------------------------------------------------------------

In a \emph{static} spacetime the standard Tolman--Ehrenfest and Tolman--Klein~\cite{TE,  Tolman-35, Klein} equilibrium conditions (see also~\cite{Tolman-30, Tolman-book}) for the locally measured temperature and chemical potential are:
\begin{equation}
T \;|| k_T || = T_\infty; 
\qquad  
\mu \;|| k_T || = \mu_\infty.
\end{equation}
Here $k_T$ is the timelike Killing vector, and in appropriate coordinates 
\begin{equation}
(k_T)^a = (\partial_t)^a = (1,0,0,0),
\end{equation} 
so that we can rephrase things as
\begin{equation}
T \; \sqrt{-g_{tt}} = T_\infty;
\qquad
\mu \; \sqrt{-g_{tt}} = \mu_\infty.
\end{equation}
(There is a minor technical assumption: we normalize $g_{tt} \to -1 $ at spatial infinity.) 
Once we add rotation life gets a little more complicated. 
The internal equilibrium conditions are now given in terms of the \emph{comoving} Killing vector
\begin{equation}
T \; || k_C || = T_\infty; 
\qquad 
\mu \; || k_C || = \mu_\infty.
\end{equation}
In appropriate coordinates we can rephrase things as
\begin{equation}
T \sqrt{-(g_{tt}+2\Omega g_{t\phi} + \Omega^2 g_{\phi\phi})} = T_\infty; \qquad
\mu \sqrt{-(g_{tt}+2\Omega g_{t\phi} + \Omega^2 g_{\phi\phi})} = \mu_\infty.
\end{equation}
Note that $T_\infty$ and $\mu_\infty$ are now defined by first going onto the rotation axis and then moving to spatial infinity. (There is a minor technical assumption that $T_\infty$ at north and south polar infinities are the same, and similarly for $\mu_\infty$.) We can also write this as
\begin{equation}
T  \sqrt{N^2-g_{\phi\phi}(\Omega-\varpi)^2} = T_\infty;
\qquad
\mu  \sqrt{N^2-g_{\phi\phi}(\Omega-\varpi)^2}= \mu_\infty.
\end{equation}
It is useful to note
\begin{equation}
||k_T|| = \sqrt{N^2-g_{\phi\phi} \varpi^2};  \qquad ||k_\Phi|| = \sqrt{g_{\phi\phi}}; \qquad  ||k_C|| = \sqrt{N^2-g_{\phi\phi}(\Omega-\varpi)^2}.
\end{equation}
In particular
\begin{equation}
||k_T|| \leq N; \qquad ||k_C|| \leq N.
\end{equation}

%------------------------------------------------------------------------------------------------------------------------------------------
\section{Entropy current}
%------------------------------------------------------------------------------------------------------------------------------------------

The entropy current (which is conserved because we are in equilibrium) is given in terms of the locally measured entropy density $s$ by 
\begin{equation}
S^a = s \; u_C^a =  s \; \left\{ { [k_T]^a + \Omega [k_\Phi]^a \over || k_T + \Omega \; k_\Phi ||} \right\}.
\end{equation}
This means the total entropy is 
\begin{equation}
S = \int s \; u_C^t  \; \sqrt{-g_4} \d^3 x =  \int s \; {1\over ||k_C||} \;  \sqrt{-g_4} \d^3 x =  \int s \; {T\over T_\infty} \;  \sqrt{-g_4} \d^3 x,
\end{equation}
where we have used the Tolman equilibrium condition.
Now apply the Euler relation 
\begin{equation}
s = {\rho+p-\mu n\over T},
\end{equation}
to obtain
\begin{equation}
S = \int   {\rho+p-\mu n\over T_\infty }\; \sqrt{-g_4} \d^3 x = {1\over T_\infty} \int   \{\rho+p-\mu n\}\; \sqrt{-g_4} \d^3 x.
\end{equation}
Similarly, the number density current is (in terms of comoving number density $n$)
\begin{equation}
j^a = n \; u_C^a =  n \; \left\{ { [k_T]^a + \Omega [k_\Phi]^a \over || k_T + \Omega \; k_\Phi ||} \right\}.
\end{equation}
This means the total number of particles is
\begin{equation}
\N = \int n \; u^t  \; \sqrt{-g_4} \d^3 x =  \int n \; {1\over ||k_C||} \;  \sqrt{-g_4} \d^3 x =  \int n \; {\mu\over \mu_\infty} \;  \sqrt{-g_4} \d^3 x,
\end{equation}
where we have used the Tolman--Klein equilibrium condition. That is 
\begin{equation}
\N = 
{1\over\mu_\infty} \int n \, {\mu} \, \sqrt{-g_4} \d^3 x,
\end{equation}
whence
\begin{equation}
S = {1\over T_\infty} \int   \{\rho+p\}\sqrt{-g_4} \; \d^3 x - {\mu_\infty \N\over T_\infty}.
\end{equation}
This is the same fundamental equation as we had for the non-rotating case~\cite{Abreu1, Abreu2} --- but the logic flow we used to get to it has been rather different. 
Assuming a non-negative chemical potential we still have
\begin{equation}
S \leq  {1\over T_\infty} \int   \{\rho+p\} \; \sqrt{-g_4} \d^3 x.
\end{equation}
As long as pressure is positive 
\begin{equation}
S \leq  {1\over T_\infty} \int   \{\rho+3 p\} \; \sqrt{-g_4} \d^3 x.
\end{equation}
This is the same fundamental inequality as we had for the non-rotating case~\cite{Abreu1, Abreu2} --- but the interpretation will now  be rather different. In the static case the quantity 
\begin{equation}
Q_S \equiv  \int   \{\rho+3 p\}\; \sqrt{-g_4} \d^3 x,
\end{equation}
is equal to the so-called Tolman mass~\cite{Tolman-30,Tolman-book}. In a rotating system this is no longer true. $Q_S$ is closely related to the Tolman mass but no longer equal to it.  This is not a problem for us, as it is this quantity $Q_S$ that we shall now bound,  and so use to produce a bound on the entropy. At this stage of the calculation we must be content with
\begin{equation}
S \leq  {Q_S\over T_\infty}.
\end{equation}

%------------------------------------------------------------------------------------------------------------------------------------------
\section{Classical entropy bound}
%------------------------------------------------------------------------------------------------------------------------------------------

To bound the quantity $Q_S$ it is useful to consider the two natural congruences on the spacetime. 
\begin{itemize}
\item 
For the comoving congruence $u_C$ we have  $u_C\propto k_T + \Omega \, k_\Phi$, so that we are dealing with a Killing congruence. 
\item 
The  second natural congruence to consider is the congruence defined by  the  FIDOs (fiducial observers), sometimes called ZAMOs (zero angular momentum observers). See for example~\cite{Membrane}. This FIDO/ZAMO congruence is specified by $u_F= -(\d t)^\#/||\d t||$, or more explicitly $(u_F)^a = - \nabla^a t / ||\nabla t||$. 
\end{itemize}
Note that in stationary axial symmetry we have
\begin{equation}
[u_C]^a = {(1,\Omega,0,0)\over ||k_C||},
\end{equation}
while
\begin{equation}
[u_F]^a = {(1,\varpi,0,0)\over N},
\end{equation}
and so in particular
\begin{equation}
\nabla \cdot u_F = 0.
\end{equation}
Since $u_C$ and $u_F$ are both timelike we know 
\begin{equation}
(u_C \cdot u_F)^2 \geq 1.
\end{equation}
We now have 
\begin{eqnarray}
Q_S &=& \int   \{\rho+3 p\}\sqrt{-g_4} \d^3 x
\\
&\leq& \int   \{ 2 \rho (u_C \cdot u_F)^2 + (-\rho+3p) \}\sqrt{-g_4} \d^3 x
\\
&\leq& \int   \{ 2 T_{ab} u_F^a u_F^b + T \}\sqrt{-g_4} \d^3 x
\\
&=&  2\int   \left\{ T_{ab} - {1\over2} T g_{ab} \right\} u_F^a u_F^b  \; \sqrt{-g_4} \d^3 x
\\
&=& {1\over4\pi} \int   \left\{ R_{ab} \right\} u_F^a u_F^b  \; \sqrt{-g_4} \d^3 x,
\end{eqnarray}
where in the last step we have used the Einstein equations. 
Now by construction the congruence $u_F$ is irrotational (vorticity free, $\omega=0$), and in addition we have seen that it is divergence free, $\theta=0$.  
The congruence $u_F$ is however not a geodesic congruence, and we let $a_F$ be the 4-acceleration of $u_F$.  It is now a standard result that the (non-geodesic) timelike Raychaudhuri equation (see appendix) implies
\begin{equation}
R_{ab} \; u_F^a u_F^b  = - \sigma^2 + \nabla\cdot a_F,
\end{equation}
where $\sigma$ is the shear of $u_F$. 
This implies
\begin{eqnarray}
Q_S \leq {1\over4\pi} \int   \left\{ - \sigma^2 + \nabla\cdot a_F \right\}  \; \sqrt{-g_4} \d^3 x
\end{eqnarray}
But we always have $\sigma^2\geq 0$, so we see
\begin{eqnarray}
Q_S &\leq& {1\over4\pi} \int   \left\{  \nabla\cdot a_F \right\}  \; \sqrt{-g_4} \d^3 x
\\
&=& {1\over4\pi} \int   \left\{  \partial_i ( \sqrt{-g_4} a_F^i ) \right\}  \d^3 x
\\
&=&  {1\over4\pi} \int   \left\{  \partial_i ( \sqrt{g_3}  N a_F^i ) \right\}  \d^3 x
\\
&=&  {1\over4\pi} \int   \left\{  N \, a_F^i  \right\}  \hat n_i \sqrt{g_2} \; \d^2 x.
\end{eqnarray}
The index $i$ runs over $r$, $\theta$, $\phi$, since any $t$ dependence is automatically eliminated by stationarity.
Now define a 3-dimensional vector
\begin{equation}
\vec \kappa = N \; \vec a_F,
\end{equation}
in terms of which we have
\begin{equation}
Q_S \leq  {1\over 4 \pi} \int   \left\{  \vec \kappa  \cdot \hat n \right\} \sqrt{g_2} \; \d^2 x.
\end{equation}
In terms of the area, $\d\A = \sqrt{g_2}\; \d^2 x$, this implies
\begin{equation}
Q_S \leq  {1\over 4 \pi} \int   ||\vec \kappa|| \; \d\A.
\end{equation}
Note that $\vec\kappa$ is a natural generalization of the usual surface gravity~\cite{Abreu1, Abreu2}, which now extends into the bulk of any general stationary axisymmetric spacetime. (And so in particular $\vec \kappa$ makes sense for the 2-surface of any arbitrary 3-volume, regardless of whether or not that 2-surface is null.) It is interesting to note that there are a large number of situations for which similar ``bulk'' extensions of the usual notion of surface gravity are important~\cite{alex, minimal, pseudo}.

\paragraph{Summarizing:} Up to this stage of the argument, purely on classical grounds, (classical general relativity plus basic thermodynamics), we see that the entropy of equilibrium uncollapsed matter confined to a region of surface area $\A$ is bounded by
\begin{equation}
\label{E:classical}
S \leq {Q_S\over T_\infty} \leq {1\over 4 \pi \,T_\infty} \int   ||\vec \kappa|| \; \d\A \leq {||\vec \kappa||_{\mathrm{max(surface)}} \; \A \over 4\pi \,T_\infty}.
\end{equation}
No appeal to quantum physics has yet been made. Note that terms being neglected in deriving this classical bound, pressures and chemical potentials, are typically smaller than or of the same order as the terms being retained. This this purely classical bound is  typically a reasonably tight  quantitative bound on the entropy.

%----------------------------------------------------
\section{Unruh temperature}
%----------------------------------------------------

We now invoke the only bit of quantum physics that enters our argument: The Unruh effect~\cite{Unruh}. This effect has now been studied for some 35 years and is closely related to the Hawking radiation effect~\cite{Hawking1, Hawking2}. Like Hawking radiation, despite some 35 years of intense theoretical effort there has as yet been no fully convincing experimental proof of the reality of this effect --- though the situation may now be changing~\cite{Silke, Belgiorno1, Belgiorno2, Belgiorno3}.  Nevertheless, the Unruh effect is based on such basic and fundamental aspects of special relativistic quantum field theory that it is extremely difficult to see how to avoid this effect without at the same time undermining many highly successful aspects of quantum field theory. Accordingly, while the Unruh effect may not have the direct experimental support of the various ingredients that went into the classical bound derived above, the existence of a quantum-induced Unruh temperature is felt (by almost everyone in the community) to be an entirely uncontroversial and plausible assumption.

Explicitly introducing the Boltzmann constant and Planck constant, for a FIDO with 4-velocity $u_F$ and 4-acceleration $a_F$ the locally measured Unruh temperature is
\begin{equation}
k_B \, T_{U,F} = {\hbar \; ||a_F||\over2\pi}.
\end{equation}
To convert this to a temperature as seen by a comoving observer $u_C$ the standard technique is to define a temperature \emph{4-vector}, $T^a = T \; u_F^a$, and perform a Lorentz transformation~\cite{Tolman-book}.  The relevant quantity is boosted by the gamma factor
\begin{equation}
\gamma = |u_F \cdot u_C| =   {|\nabla t \cdot (k_T+\Omega \; k_\Phi)| \over ||\nabla t|| \; ||k_C||} =  {1\over (1/N) \; ||k_C||} = {N\over ||k_C||} \geq 1.
\end{equation}
Thus the effective Unruh temperature associated with the FIDO $u_F$, as seen by a comoving observer $u_C$, is
\begin{equation}
k_B T_{U,C} = {\hbar \; ||a_F||  \; N \over2\pi  || k_C||}.
\end{equation}
Now, as seen from infinity along the axis of rotation we have seen that the relevant redshift factor (based on the Tolman--Ehrenfest relation~\cite{TE, Tolman-35}) is
\begin{equation}
T_{U,\infty} = T_{U,C} \;  || k_C||.
\end{equation}
So finally the Unruh temperature associated with the FIDO $u_F$,  as seen from spatial infinity, is
\begin{equation}
k_B \; T_{U,\infty} =  {\hbar \; ||a_F||\; N \over2\pi}  =  {\hbar \; ||\vec \kappa||  \over2\pi} .
\end{equation}
This implies that in any physical equilibrium system the Unruh effect can be used to argue for an ultimate and universal lower bound on the equilibrium temperature
\begin{equation}
k_B T_{\infty}  \geq \max\nolimits_{_\mathrm{FIDOs}} \left\{ {\hbar \; ||\vec\kappa || \over2\pi} \right\} .
\end{equation}
We have explicitly included the $\hbar$ to make it absolutely clear that this appeal to the Unruh effect~\cite{Unruh} is the only part of the argument that in any way involves quantum physics.  In more traditional theorists' units we can write this as
\begin{equation}
\label{E:temperature}
T_{\infty}  \geq \max\nolimits_{_\mathrm{FIDOs}} \left\{ {||\vec\kappa || \over2\pi} \right\} .
\end{equation}
Note that while formally this result is identical to the static spacetime result reported in~\cite{Abreu1, Abreu2}, the technical issues underlying the argument are now considerably more subtle.  Also note that this quantum-inspired semiclassical bound, while extremely general, is also often quantitatively weak:  For objects such as stars or even planets, the actual temperature is often very many orders of magnitude higher than this quantum-inspired bound. Because of this, the resulting semiclassical entropy bound is sometimes rather weak --- quantitatively weak but qualitatively robust.  To get a tight quantitative bound on the temperature (and thus the entropy) we would need to consider some ultra-compact object (possibly a monster~\cite{monster1, monster2, monster3}/ gravastar~\cite{gravastar1, gravastar2, gravastar3, gravastar4}, black star~\cite{black-star1, black-star2, black-star3, black-star4}, or quasi-black hole~\cite{qbh1, qbh2}) whose actual temperature was close to the Hawking temperature it would have if it were to collapse to a black hole.

%----------------------------------------------------
\section{The semiclassical bound}
%----------------------------------------------------

Having done all the preparatory work we note
\begin{equation}
S \leq  {||\vec \kappa||_\mathrm{max(surface)} \; \A \over 4\pi \,T_\infty};   
\qquad
T_{\infty}  \geq  {||\vec\kappa ||_\mathrm{max(FIDOs)} \over2\pi}.
\end{equation}
Therefore
\begin{equation}
S \leq \left\{ {||\vec \kappa||_\mathrm{max(surface)}  \over ||\vec\kappa ||_\mathrm{max(FIDOs)} }\right\} \; {\A\over 2}.
\end{equation}
In particular, in the numerator we are maximizing only over those FIDOs that skim the surface of the object, while in the denominator we are maximizing over the larger class of all FIDOs in the bulk, therefore this bracketed ratio is less than or equal to unity.  
Finally, as claimed,
\begin{equation}
S \leq {\A\over 2}.
\end{equation}
The particularly nice feature of this bound is how general it is and how weak the assumptions are that go into it. 
There are a number of places where the  use of inequalities has been sub-optimal, particularly when it comes to the semiclassical temperature bound, and in situations where one knows more about the internal structure of the region enclosed by the surface of area $\A$ one might potentially be able to obtain tighter results. (See for example the discussion in reference~\cite{Abreu1}.) However in general the ${1\over2}$ seems to be an intrinsic feature for uncollapsed matter, ultimately arising from the use of the Euler relation, which in turn ultimately depends on temperature being intensive and entropy being extensive~\cite{Abreu1}.  

In counterpoint, note that for matter that has collapsed to a black hole one does \emph{not} have the usual Euler relation. For example, for Schwarzschild black holes in standard general relativity one has $T\propto 1/M$ and $S\propto M^2$. So temperature is no longer intensive and entropy is no longer extensive. The closest equivalent to the usual Euler relation for uncollapsed matter
\begin{equation}
\rho =  T s + \mu n - p,
\end{equation}
is now (for collapsed matter) the Smarr mass formula for standard general relativity black holes~\cite{Smarr, mechanix}
\begin{equation}
M = 2 \,T \,S + 2 \,\Omega \,J + Q \,\Phi_H.
\end{equation}
The key point here is the relative factor $2$ between these two equations, which ultimately leads to   (entropy) $\leq {1\over2}$ (area) for uncollapsed matter.

%----------------------------------------------------
\section{Discussion}
%----------------------------------------------------

While we suspect that there might still be a number of ways in which (in specific situations) the bounds enunciated in this article can be improved, the overall message is (we think) clear:  Useful (albeit sometimes quantitatively weak) entropy bounds can be derived from very basic physics without any recourse to the long sought for ``full theory of quantum gravity''. Our first bound, summarized  in equation (\ref{E:classical}),  was purely classical --- using only classical general relativity and basic thermodynamics to place a (quantitatively strong)  upper bound on the entropy.  A second bound, summarized in equation (\ref{E:temperature}), is semiclassical and appealed to the Unruh effect (the only quantum physics involved in our argument) to place a (quantitatively weak but qualitatively robust) lower bound on the temperature as seen at infinity.  Combining these two bounds then yields  our final result: (entropy) $\leq {1\over2}$ (area) for uncollapsed matter.  These are remarkably useful bounds based on an absolute minimum of physical assumptions.

%-------------------------------------------------
\appendix
%%-----------------------------------------------
\section{The non-geodesic timelike Raychaudhuri equation}
%%-----------------------------------------------

Let $u^a$ be a field of unit timelike vectors (a congruence).  This does not have to be the 4-velocity of a physical fluid (though it might be), it applies just as well to the 4-velocities of an imaginary collection of ``fiducial observers'' [FIDOs]. As is completely standard, (give or take the odd factor of 2 in the definitions of $\sigma^2$ and $\omega^2$), let us define~\cite{Hawking-Ellis, Wald}
\begin{equation}
h_{ab} = g_{ab} + u_a u_b; 
\qquad
\theta_{ab} = h_{ac} \nabla^{(c} u^{d)} h_{db};
\qquad
\theta=  g^{ab}\theta_{ab} = h^{ab} \theta_{ab} = \nabla_a u^a;
\end{equation}
\begin{equation}
\sigma_{ab} = \theta_{ab} - {1\over3} h_{ab} \theta;
\qquad
\sigma^2 = \sigma_{ab} \sigma^{ab} \geq 0;
\end{equation}
and
\begin{equation}
\omega_{ab} = h_{ac} \nabla^{[c} u^{d]} h_{db};
\qquad
\omega^2 =  \omega_{ab} \omega^{ab} \geq 0.
\end{equation}
With these definitions we have the decomposition~\cite{Hawking-Ellis, Wald}
\begin{equation}
u_{a;b} = \omega_{ab} + \sigma_{ab} +{1\over3} \theta h_{ab} - {\d u_a\over\d s} \; u_b.
\end{equation}
Then it is a purely geometrical result (see for example Hawking and Ellis~\cite{Hawking-Ellis}, pages 82--84, or Wald~\cite{Wald} page 218) that
\begin{equation}
{\d\theta\over\d s} = - R_{ab} u^a u^b +  \omega^2 -  \sigma^2 - {1\over3}\theta^2 + \nabla_a \left( {\d u^a\over\d s} \right).
\end{equation}
\emph{This is the standard form of the (non-geodesic) Raychaudhuri equation.}
This is Wald's (9.2.11), supplemented with the $\nabla_a \left( {\d u^a\over\d s} \right)$ term due to a non-geodesic congruence, the presence of which you can deduce from the second line in his (9.2.10) by not assuming geodesic motion.
Note also
\begin{equation}
{\d \theta\over \d t} = u \cdot \nabla \theta = \nabla\cdot (\theta u) - \theta \nabla\cdot u =  \nabla\cdot (\theta u) - \theta^2,
\end{equation}
so that we can also write the Raychaudhuri equation in the slightly unusual forms
\begin{equation}
\nabla_a \left( \theta u^a - {\d u^a\over\d s} \right)= - R_{ab} u^a u^b +  \omega^2 - \sigma^2 + {2\over3}\theta^2,
\end{equation}
or
\begin{equation}
R_{ab} \, u^a u^b= +  \omega^2 -  \sigma^2 + {2\over3}\theta^2 +  \nabla_a \left(- \theta \; u^a + {\d u^a\over\d s} \right).
\end{equation}
For the particular case of the FIDO congruence $u_F = -(\d t)^\sharp/||\d t||$ we automatically have $\omega=0$, and from stationarity plus axisymmetry we have $\theta=0$, thus in this situation the non-geodesic Raychaudhuri equation reduces (as claimed) to
\begin{equation}
R_{ab} \, u_F^a u_F^b=  -  \sigma^2  + \nabla_a  (a_F)^a.
\end{equation}

%-----------------------------------------------
\section*{Acknowledgements}
%-----------------------------------------------

This research was supported by the Marsden Fund administered by the Royal Society of New Zealand. 

%-------------------------------

%-------------------------------
%----------------------------------------------------
\end{document}